\documentclass[conference]{IEEEtran}
\IEEEoverridecommandlockouts
\usepackage{graphics} 
\usepackage{graphicx} 
\usepackage{epsfig} 
\usepackage{times} 
\usepackage{amsmath} 
\usepackage{amssymb}  
\usepackage{array}
\usepackage{cite}
\usepackage{CJK}
\usepackage{color}
\usepackage{comment}
\usepackage{epstopdf}
\usepackage{subfigure}
\usepackage{float} 
\usepackage{verbatim}
\usepackage{tabularx}
\usepackage{booktabs} 
\usepackage{multirow}
\usepackage{mdwmath}
\usepackage{mdwtab}
\usepackage{epsfig}
\usepackage{eqparbox}
\usepackage{graphicx}
\usepackage{diagbox}
\usepackage{siunitx}
\usepackage{mathrsfs}
\usepackage{amsmath}
\usepackage{amssymb}
\usepackage{xcolor}
\usepackage{makecell}

\usepackage[ruled,linesnumbered]{algorithm2e}
\usepackage{url}
\newcommand{\method}[1]{\texttt{#1}}

\makeatletter
\let\NAT@parse\undefined
\makeatother
\usepackage[
colorlinks=true,      
linkcolor=black,      
citecolor=black,      
filecolor=black,      
urlcolor=black]{hyperref}

\newtheorem{definition}{Definition}

\newtheorem{remark}{Remark}
\newtheorem{lemma}{Lemma}

\def\BibTeX{{\rm B\kern-.05em{\sc i\kern-.025em b}\kern-.08em
    T\kern-.1667em\lower.7ex\hbox{E}\kern-.125emX}}
\begin{document}

\title{Robust Data-EnablEd Predictive Leading Cruise Control via Reachability Analysis\\
\thanks{This work is supported by National Natural Science Foundation of China under grant 52302410, Postdoctoral Fellowship Program of CPSF under grant GZB20230354, Tsinghua University-Huawei Joint Research Project, Young Elite Scientists Sponsorship Program by CHINA-SAE, and Shuimu Tsinghua Scholarship. Corresponding author: Jiawei Wang and Keqiang Li.}
}

\author{\IEEEauthorblockN{1\textsuperscript{st} Shuai Li}
\IEEEauthorblockA{\textit{School of Vehicle and Mobility} \\
\textit{Tsinghua University}\\
Beijing, China \\
li-s21@mails.tsinghua.edu.cn}\\

\vspace{-1mm}
\IEEEauthorblockN{4\textsuperscript{th} Jiawei Wang}
\IEEEauthorblockA{\textit{School of Vehicle and Mobility} \\
\textit{Tsinghua University}\\
Beijing, China \\
wang-jw18@tsinghua.org.cn}\\

\and

\IEEEauthorblockN{2\textsuperscript{nd} Chaoyi Chen}
\IEEEauthorblockA{\textit{School of Vehicle and Mobility} \\
\textit{Tsinghua University}\\
Beijing, China \\
chency2023@tsinghua.edu.cn}\\

\vspace{-1mm}
\IEEEauthorblockN{5\textsuperscript{th} Qing Xu}
\IEEEauthorblockA{\textit{School of Vehicle and Mobility} \\
\textit{Tsinghua University}\\
Beijing, China \\
qingxu@tsinghua.edu.cn}\\

\and

\IEEEauthorblockN{3\textsuperscript{rd} Haotian Zheng}
\IEEEauthorblockA{\textit{School of Vehicle and Mobility} \\
\textit{Tsinghua University}\\
Beijing, China \\
zhenght21@mails.tsinghua.edu.cn}\\

\vspace{-1mm}
\IEEEauthorblockN{6\textsuperscript{th} Keqiang Li}
\IEEEauthorblockA{\textit{School of Vehicle and Mobility} \\
\textit{Tsinghua University}\\
Beijing, China \\
likq@tsinghua.edu.cn}\\
}

\maketitle

\begin{abstract}
Data-driven predictive control promises model-free wave-dampening strategies for Connected and Autonomous Vehicles (CAVs) in mixed traffic flow. However, its performance relies on data quality, which suffers from unknown noise and disturbances. 
This paper introduces a Robust Data-EnablEd Predictive Leading Cruise Control (\method{RDeeP-LCC}) method based on reachability analysis, aiming to achieve safe and optimal CAV control under bounded process noise and external disturbances. Precisely, the matrix zonotope set technique and Willems' Fundamental Lemma are employed to derive the over-approximated system dynamics directly from data, and a data-driven feedback control technique is utilized to obtain an additional feedback input for stability. We decouple the mixed platoon into an error system and a nominal system, where the error system provides data-driven reachability sets for the enhanced safety constraints in the nominal system. Finally, a data-driven predictive control framework is formulated in a tube-based control manner for robustness guarantees. Nonlinear simulations with noise-corrupted data demonstrate that the proposed method outperforms baseline methods in mitigating traffic waves.
\end{abstract}


\section{Introduction}
Platooning technologies involving multiple Connected and Autonomous Vehicles (CAVs) have experienced rapid development~\cite{guanetti2018control}. By exploiting the benefits of vehicle-to-vehicle communications and cooperative control algorithms, significant improvements have been achieved in traffic safety, energy economy, and driving comfort for pure CAV  platoons~\cite{hu2022distributed}. However, the full marketization of CAVs is an evolving process. A mixed traffic scenario will persist during this transitional period, characterized by the coexistence of CAVs and Human-Driven Vehicles (HDVs)~\cite{wang2020controllability,li2023information}. 

Unlike pure CAV platoons, mixed platoons incorporating CAVs and HDVs do not require all vehicles to have autonomous driving capabilities, making them directly applicable to mixed traffic. By appropriate design of CAV cooperation strategies, recent research has revealed the potential of mixed platoons in attenuating undesired traffic perturbations and smoothing traffic flow~\cite{wu2021flow,wang2021leading}. Existing research mostly captures the  HDVs' dynamics by a car-following model and employs model-based control methods for the CAVs. These methods include optimal control~\cite{jin2016optimal}, Model Predictive Control (MPC)~\cite{Feng2021robust}, and control barrier function~\cite{zhao2023safety}. However, their performance relies heavily on the model accuracy, while in practice, the HDVs' dynamics are usually unknown and uncertain. In contrast, model-free methods, which bypass the need for prior knowledge of mixed platoon dynamics, have recently shown their potential in learning CAV control policies from data, as evidenced by techniques such as adaptive dynamic programming~\cite{huang2020learning} and reinforcement learning~\cite{wu2021flow}. Nevertheless, these methods typically require iterative approximations to find optimal solutions, which can significantly increase computational demands. Also, their lack of guaranteed safety and interpretability limits their practical implementations. 

On the other hand, combining data-driven methods and the well-established MPC, data-driven predictive control promises safe and optimal controllers from data~\cite{hewing2020learning}. One notable technique is Data-EnablEd Predictive Control (DeePC)~\cite{Coulson2019data}, which utilizes Willems' fundamental lemma~\cite{Willems2005note} to represent the behavior of unknown systems in a data-centric manner. By adapting DeePC to a Leading Cruise Control (LCC) framework~\cite{wang2021leading}, the recently introduced Data-EnablEd Predictive Leading Cruise Control (\method{DeeP-LCC}) allows for CAVs' optimal cooperation in mixed platoons from measurable traffic data with explicit safety constraints~\cite{Wang2023deep}. Both nonlinear traffic simulations~\cite{Wang2023deep} and real-world experiments~\cite{wang2023implementation} have validated its wave-dampening capability without requiring knowledge of surrounding HDVs' car-following dynamics. However, existing research has overlooked the noise impact on offline data collection and online predictive control, and imposes over-simplified assumptions for external disturbances, which could still raise safety concerns~\cite{zhao2023robust}. Indeed, emerging evidence has suggested that robustifying standard DeePC could significantly enhance its safety performance~\cite{Huang2023robust,Berberich2020data}. Along this direction, a very recent paper has reformulated \method{DeeP-LCC} by min-max robust optimization to tackle unknown disturbances on the head vehicle~\cite{Shang2023smoothing}, but the noise issue for the overall traffic has not been well addressed. 

Compared to min-max reformulation, reachability analysis provides a computationally more reliable technique for robustness against all possible noise and disturbances. Recent studies have utilized this technique to design robust control strategies for CAVs; see, \emph{e.g.}, safety controller synthesis via backward reachability analysis~\cite{schurmann2021formal} and anti-adversary control using reach-avoid specification~\cite{xu2022reachability}. Note that these works are mostly model-based, with one very recent exception in~\cite{lan2021data}, whose prediction accuracy, however, is constrained due to the absence of a data-driven dynamics model.

To address the aforementioned issues, this paper proposes a Robust Data-EnablEd Predictive Leading Cruise Control (\method{RDeeP-LCC}) method via data-driven reachability analysis, aiming to design robust control strategies for CAVs against unknown noise and disturbances. Our contributions include: 1) We decouple the original mixed platoon system into an idealized nominal system and an error system with bounded noise and disturbances. Inspired by~\cite{lan2021data,alanwar2023data}, the data-driven reachable sets are calculated for the error system by modeling the system dynamics as an over-approximated matrix zonotope set, and an additional data-driven feedback control law is designed for stability. Similarly to a tube-based MPC mechanism, the reachable sets of the error system are utilized to tighten the safety constraints for the nominal system. Then, a data-driven predictive control formulation is proposed with this enhanced constraint, which, combined with the feedback control law of the error system, provides the CAVs' control inputs with robustness guarantees. 2) Nonlinear simulations with noise-corrupted data and disturbances validate the performance of \method{RDeeP-LCC} in mitigating traffic waves and tracking the equilibrium state. In a series of $20$ tests with randomly generated offline datasets, \method{RDeeP-LCC} consistently demonstrates superior robustness compared to the standard \method {DeeP-LCC}~\cite{Wang2023deep} or MPC with prior dynamics knowledge. 

The rest of this paper is organized as follows. Section~\ref{Sec:2} provides preliminaries and system modeling. Section~\ref{Sec:3} introduces the \method{RDeeP-LCC} method. Section~\ref{Sec:4} shows numerical simulations, and Section~\ref{Sec:5} concludes this paper.

\section{Preliminaries and System Modeling}
\label{Sec:2}

In this section, we present some preliminaries and the parametric modeling of the mixed platoon system.

\subsection{Preliminaries}

\begin{definition}[Zonotope Set~\cite{Althoff2010reachability}]
	Given a center vector $c_\mathcal{Z} \in \mathbb{R} ^n$, and $\gamma_\mathcal{Z} \in \mathbb{N}$ generator vectors in a generator matrix $G_\mathcal{Z} =\left[g_\mathcal{Z}^{(1)}, g_\mathcal{Z}^{(2)},\ldots, g_\mathcal{Z}^{(\gamma_\mathcal{Z})}\right] \in \mathbb{R}^{n \times \gamma_\mathcal{Z}}$, a zonotope set is defined as $\mathcal{Z} =\left \langle c_\mathcal{Z}, G_\mathcal{Z}\right \rangle = \left\{x \in \mathbb{R}^n \mid x=c_\mathcal{Z}+\sum_{i=1}^{\gamma_\mathcal{Z}} \beta^{(i)} g_\mathcal{Z}^{(i)},-1 \leq \beta^{(i)} \leq 1\right\}$. For zonotope sets, the following operations hold:
	\begin{itemize}
		\item  \textit{Linear Map:} For a zonotope set $\mathcal{Z}=\left \langle c_\mathcal{Z}, G_\mathcal{Z}\right \rangle $, $L \in \mathbb{R}^{m\times n}$, the linear map is defined as $L\mathcal{Z}=\left \langle Lc_\mathcal{Z}, LG_\mathcal{Z}\right \rangle$.
		
		\item  \textit{Minkowski Sum:} Given two zonotope sets $\mathcal{Z}_1=\left \langle c_{\mathcal{Z}_1}, G_{\mathcal{Z}_1}\right \rangle$ and $\mathcal{Z}_2=\left \langle c_{\mathcal{Z}_2}, G_{\mathcal{Z}_2}\right \rangle$ with compatible dimensions, the Minkowski sum is defined as $\mathcal{Z}_1 + \mathcal{Z}_2=\left\langle c_{\mathcal{Z}_1}+c_{\mathcal{Z}_2},\left[G_{\mathcal{Z}_1}, G_{\mathcal{Z}_2}\right]\right\rangle$. 
		
		\item  \textit{Cartesian Product:} Given two zonotope sets $\mathcal{Z}_1=\left \langle c_{\mathcal{Z}_1}, G_{\mathcal{Z}_1}\right \rangle$ and $\mathcal{Z}_2=\left \langle c_{\mathcal{Z}_2}, G_{\mathcal{Z}_2}\right \rangle$, the cartesian product is defined as
		\vspace{-0.1cm}
		\begin{equation*}
		\label{Cartesian}
		\mathcal{Z}_1 \times\mathcal{Z}_2  =\left\langle\begin{bmatrix}
		c_{\mathcal{Z}_1} \\
		c_{\mathcal{Z}_2}
		\end{bmatrix},\begin{bmatrix}
		G_{\mathcal{Z}_1} & 0 \\
		0 & G_{\mathcal{Z}_2}
		\end{bmatrix}\right\rangle.
		\end{equation*}
	\end{itemize}
\end{definition}


\begin{definition}[Matrix Zonotope Set~\cite{Althoff2010reachability}]
	Given a center matrix $C_\mathcal{M} \in \mathbb{R} ^{n \times m}$, and $\gamma_\mathcal{M} \in \mathbb{N}$ generator matrices in a generator matrix $G_\mathcal{M} =\left[g_\mathcal{M}^{(1)}, g_\mathcal{M}^{(2)},\ldots, g_\mathcal{M}^{(\gamma_\mathcal{M})}\right] \in \mathbb{R}^{n \times m\gamma_\mathcal{M}}$, a matrix zonotope set is defined as $\mathcal{M} =\left \langle C_\mathcal{M}, G_\mathcal{M}\right \rangle = \left\{X \in \mathbb{R}^{n \times m} \mid X=C_\mathcal{M}+\sum_{i=1}^{\gamma_\mathcal{M}} \beta^{(i)} G_\mathcal{M}^{(i)},-1 \leq \beta^{(i)} \leq 1\right\}$.
\end{definition}


\begin{definition}[Persistently excitation~\cite{Willems2005note}]
	\label{Definition:HankelMatrix}
	Given a signal sequence $\omega  = \mathrm{col}(\omega (1),\omega (2),\ldots,\omega (T))$ of length $T \in \mathbb{N}$~\footnote{Given vectors or matrices $X_0,X_1,\ldots,X_n$ with compatible sizes, we denote $\mathrm{col}(X_0,X_1,\ldots,X_n)=[X^{\top}_0,X^{\top}_1,\ldots,X^{\top}_n]^{\top}$.}, the sequence $ \omega $ is persistently exciting with order $l\in \mathbb{N}$ if and only if the following Hankel matrix is of full row rank:
	\begin{equation}
	\label{HankelMatrix}
	\mathcal{H}_l(\omega)=\begin{bmatrix}
	\omega(1) & \omega(2)   & \cdots & \omega(T-l+1) \\
	\omega(2) & \omega(3)   & \cdots & \omega(T-l+2) \\
	\vdots    & \vdots      & \ddots & \vdots \\
	\omega(l) & \omega(l+1) & \cdots & \omega(T)
	\end{bmatrix}.    
	\end{equation}
\end{definition}


\begin{lemma}[Willems' Fundamental Lemma~\cite{Willems2005note}]      
	\label{Lemma:Fundamental}
	Consider a controllable Linear Time-Invariant (LTI) system. Let $u^{\rm{d}}=\mathrm{col}(u(1),u(2),\ldots,u(T))$ be an input sequence persistently exciting with order $L+n$, where $n$ is the dimension of the system state, and the corresponding state sequence is $x^{\rm{d}}=\mathrm{col}(x(1),x(2),\ldots,x(T))$. Then $u^{\rm{s}}$ and $x^{\rm{s}}$ is a length-\textit{L} input–output trajectory of the system if and only if there exists a vector $g \in \mathbb{R}^{T-L+1}$ satisfying
	\begin{equation}
	\label{Eq:Williems}
	\begin{bmatrix}
	\mathcal{H}_L\left(u^{\rm{d}}\right) \\
	\mathcal{H}_L\left(x^{\rm{d}}\right)
	\end{bmatrix} g=\begin{bmatrix}
	u^{\rm{s}} \\
	x^{\rm{s}}
	\end{bmatrix}.
	\end{equation}
\end{lemma}

The physical interpretation of Lemma~\ref{Lemma:Fundamental} is that for a controllable LTI system, the subspace consisting of all feasible trajectories ($u^{\rm{s}},x^{\rm{s}}$) of length $L$ is identical to the space spanned by a Hankel matrix of order $L$ constructed from pre-collected data ($u^{\rm{d}},x^{\rm{d}}$) with rich enough control inputs.

\subsection{Parametric Modeling of Mixed Platoon System}

\begin{figure}[t]
	\vspace{1mm}
	\centering
	{\includegraphics[width=8.6cm]{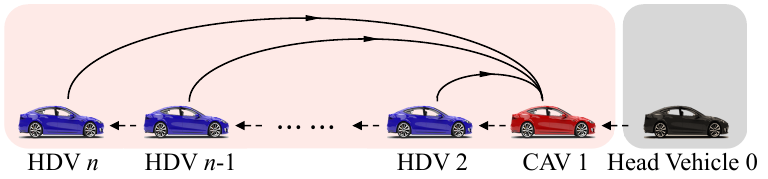}}\\
	\vspace{-2mm}
	\caption{Schematic for a CF-LCC mixed platoon. The platoon consists of one leading CAV (colored in red) and multiple following HDVs (colored in blue), with one head vehicle (colored in black) at the very beginning.}
	\label{Fig:MixedPlatoon}
	\vspace{-3mm}
\end{figure}

As depicted in Fig.~\ref{Fig:MixedPlatoon}, we analyze a mixed platoon system comprising one head vehicle (indexed as $0$), one leading CAV (indexed as $ 1 $), and $ n-1$ following HDVs (indexed as $ 2, \ldots,n$ against the moving direction). This kind of system is called as Car-Following Leading Cruise Control (CF-LCC) in~\cite{wang2021leading}, and practical mixed traffic can be naturally partitioned into several CF-LCC systems. Define $ \Omega =\{0, 1, 2, \ldots, n\}  $ as the vehicles' index set. 

In the following, we present the common parametric modeling process for the mixed platoons. For all the vehicles $ i \in \Omega $, a typical second-order linear longitudinal dynamics~\cite{wang2020controllability,jin2016optimal} are utilized, given by:
\begin{equation}
\label{Eq:DynamicsModel}
\begin{cases}
\dot{p}_i(t) = v_i(t), \\
\dot{v}_i(t) = u_i(t), \\
\end{cases}
\end{equation}
where $u_i(t)$ represents the control input, and $p_i(t)$ and $v_i(t)$ represent the position and velocity, respectively. 

For HDVs, the control input $u_i(t)$ is determined by the drivers, and can be represented by well-established car-following models like Optimal Velocity Model (OVM)~\cite{Bando1995dynamical}, with the general form given by:
\begin{equation}
\label{Eq:HDV_General}
u_i(t) = F(s_i(t),\dot{s}_i(t),v_i(t)),
\end{equation}
where $s_i(t) = p_{i-1}(t) - p_i(t)$ and $\dot{s}_i(t) = v_{i-1}(t) - v_i(t)$ denote the spacing and relative velocity between vehicle $i$ and its preceding vehicle, respectively. For simplicity, a homogeneous expression of $F$ is utilized for parametric modeling in this section, but the proposed data-driven control method in this paper is directly applicable to general heterogeneous cases.

Define $\tilde{s}_i(t)$ and $\tilde{v}_i(t)$ as the deviations of spacing and velocity from the equilibrium state, respectively:
\begin{equation}
\label{Eq:Equilibrium}
\tilde{s}_i(t)={s}_i(t)-s^{*},	 \quad
\tilde{v}_i(t)={v}_i(t)-v^{*},
\end{equation}
where $s^*$ and $v^*$ denote the spacing and velocity in traffic equilibrium, satisfying the condition $F(s^*,0,v^*) = 0$.

Then, the first-order Taylor expansion can be applied to~\eqref{Eq:HDV_General}, which, combined with~\eqref{Eq:DynamicsModel}, yields the HDVs' linearized model:
\begin{equation}
\label{Eq:HDV_Linear}
\begin{cases} 
\dot{\tilde{s}}_i(t)=\tilde{v}_{i-1}(t) -\tilde{v}_i(t), \\
\dot{\tilde{v}}_i(t)=\gamma_{1} \tilde{s}_{i}(t)+\gamma_{2} \tilde{v}_{i}(t)+\gamma_{3} \tilde{v}_{i-1}(t), \\
\end{cases}
\end{equation}
where $\gamma_{1}=\frac{\partial F}{\partial s}$, $\gamma_{2}=\frac{\partial F}{\partial \dot{s}} - \frac{\partial F}{\partial v}$, and $\gamma_{3}=\frac{\partial F}{\partial \dot{s}}$ are evaluated at the equilibrium state $(s^*, v^*)$. 

For the CAV, its linearized longitudinal dynamics can be given in the following form:
\begin{equation}
\label{Eq:DynamicsModelCAV_L}
\begin{cases}
\dot{\tilde{s}}_i(t)=\tilde{v}_{i-1}(t) -\tilde{v}_i(t), \\
\dot{\tilde{v}}_i(t)={u}_i(t),\\
\end{cases}
\end{equation}
where $u_i(t)$ is the designed control input for the CAV.

We define $x_i(t)=\left[\tilde{s}_i(t), \tilde{v}_i(t)\right]^{\top}$ as the state vector for vehicle $i$, and lump all the states of the HDVs and CAV to obtain the mixed platoon system state $x(t)= \left[x_1(t),x_2(t),\ldots,x_n(t)\right]^{\top} \in \mathbb{R}^{2n \times 1}$. Then, the state-space model of the mixed platoon is obtained as:
\begin{equation}
\label{Eq:ContinuousSystem}
\dot{x}(t)=A_{\rm{con}}x(t)+B_{\rm{con}}u(t)+H_{\rm{con}}\epsilon(t)+ w(t),
\end{equation}	  
where $A_{\rm{con}} \in \mathbb{R}^{2n \times 2n},B_{\rm{con}} \in \mathbb{R}^{2n \times 1},H_{\rm{con}} \in \mathbb{R}^{2n \times 1}$ represent the dynamics matrix, the control input matrix, and the disturbance input matrix, respectively, and $u(t) \in \mathbb{R},\epsilon(t)=\tilde{v}_0(t) \in \mathbb{R},w(t) \in \mathbb{R}^{2n \times 1}$ denote the control input of CAV, the velocity deviation of the head vehicle (external disturbance), and the unknown but bounded process noise, respectively. 

Then, one can discretize the continuous system~\eqref{Eq:ContinuousSystem} using the forward Euler method to obtain the discrete system model:
\begin{equation}
\label{Eq:DiscreteSystem}
{x}(k+1)=Ax(k)+Bu(k)+H \epsilon (k)+ w(k),
\end{equation}
where $k$ is the discrete time step, $A$, $B$, and $H$ are system matrix, control input matrix, disturbance input matrix of the discrete system, respectively.

\begin{remark}
	It is worth noting that the uncertain and unknown nature of HDVs' behaviors makes it non-trivial to accurately identify the parametric model~\eqref{Eq:DiscreteSystem},  which motivates us to develop a data-driven predictive control method. We present the general form of model~\eqref{Eq:DiscreteSystem} to facilitate our following design of the data-driven dynamics and reachability analysis, and it will also be utilized for the baseline MPC method in the simulations. Due to the page limit, interested readers are referred to~\cite{wang2021leading} for the specific expressions of the parametric model.
\end{remark}


\section{Methodology of \method{RDeeP-LCC} }
\label{Sec:3}
This section proposes the \method{RDeeP-LCC} method for mixed platoon control, consisting of three steps: data collection, data-driven reachable set computation, and \method{RDeeP-LCC} optimization formulation. 

\begin{figure}[t]
	\vspace{1mm}
	\centering
	{\includegraphics[width=8.6cm]{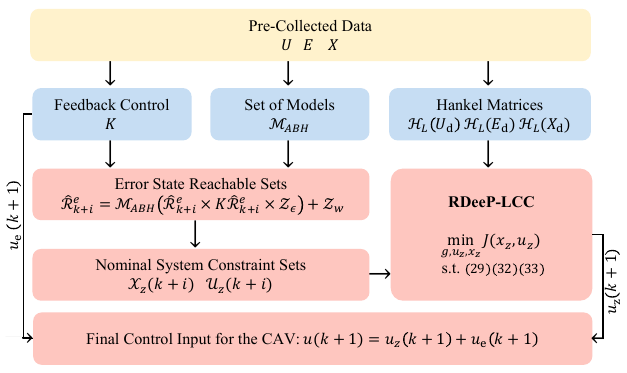}}\\
	\vspace{-2mm}
	\caption{An overview of the proposed \method{RDeeP-LCC} method.}
	\label{Fig:RDeePCMethod}
	\vspace{-3mm}
\end{figure}

Fig.~\ref{Fig:RDeePCMethod} demonstrates the flow chart of the proposed method. In the offline phase (blue), pre-collected data (yellow) is utilized to calculate the over-approximated system matrix $\mathcal{M}_{ABH}$ (blue, center), derive a probabilistic guaranteed feedback control law $K$ to ensure stability for all possible systems (blue, left), and generate the Hankel matrices (blue, right). In the online phase (pink), \method{RDeeP-LCC} solves for optimal control input for the CAV in a receding horizon manner. First, based on $\mathcal{M}_{ABH}$ and $K$, the method recursively derives the data-driven reachable sets of error states (pink, top left), which is further converted into a tightened nominal system constraint (pink, center left). Then, the nominal control input is obtained from the \method{RDeeP-LCC} optimization formulation (pink, right). Finally, the actual control input of the CAV is obtained by combining the nominal control input with the error feedback control input using the tube-based MPC mechanism to ensure robustness property (pink, bottom).

\subsection{Data Collection}
\label{Sec:3A}
For data collection, a persistently exciting sequence $ u(k) $ and  $\epsilon(k) $ of length $ T+1 $ is needed to be applied to the mixed platoon system. Specifically, define the control input sequence $U$, the disturbance input sequence $E$, and the corresponding state sequence $X$ as follows:
\begin{subequations}
	\label{Eq:Xsequence}
	\begin{equation}
	U=[u(1),u(2),\ldots,u(T+1)] \in \mathbb{R} ^{1 \times (T+1)},
	\end{equation}
	\begin{equation}
	E=[\epsilon(1),\epsilon(2),\ldots,\epsilon(T+1)] \in \mathbb{R} ^{1 \times (T+1)},
	\end{equation}
	\begin{equation}
	X=[x(1),x(2),\ldots,x(T+1)] \in \mathbb{R} ^{2n \times (T+1)}.
	\end{equation}
\end{subequations}
For the reachable set computation, the data sequences are further reorganized into a specific format:
\begin{subequations}
	\begin{equation}
	U_-=[u(1),u(2),\ldots,u(T)] \in \mathbb{R} ^{1 \times T},
	\end{equation}
	\begin{equation}
	E_-=[\epsilon(1),\epsilon(2),\ldots,\epsilon(T)] \in \mathbb{R} ^{1 \times T},
	\end{equation}
	\begin{equation}
	X_-=[x(1),x(2),\ldots,x(T)] \in \mathbb{R} ^{2n \times T},
	\end{equation}
	\begin{equation}
	X_+=[x(2),x(3),\ldots,x(T+1)] \in \mathbb{R} ^{2n \times T}.
	\end{equation}
\end{subequations}
In addition, for convenience in the subsequent derivation, we denote the sequence of unknown process noise as
\begin{equation}
W_-=[w(1),w(2),\ldots,w(T)] \in \mathbb{R} ^{2n \times T},
\end{equation}    
although it is worth noting that $W_-$ is not measurable. 

In addition, we reformulate the trajectory data $ U_- $, $ E_- $, and $ X_- $ into a compact form as the following column vectors:
\begin{subequations}
	\begin{equation}
	U_{\mathrm{d}}={\mathrm{col}}(U_-) \in \mathbb{R} ^{T},
	\end{equation}
	\begin{equation}
	E_{\mathrm{d}}={\mathrm{col}}(E_-) \in \mathbb{R} ^{T},
	\end{equation}
	\begin{equation}
	X_{\mathrm{d}}={\mathrm{col}}(X_- ) \in \mathbb{R} ^{2nT}.
	\end{equation}
\end{subequations}
Then the data $U_d$, $E_d$, $X_d$ are utilized to form the Hankel matrices by Definition~\ref{Definition:HankelMatrix}. These matrices are further partitioned into two parts, corresponding to the trajectory data in the past $ T_{\mathrm{ini}} \in \mathbb{N}$ steps and the trajectory data in the future $ N \in \mathbb{N}$ steps, defined as follows:
\begin{equation}
\label{Eq:Hankel}
\begin{bmatrix}
U_{\mathrm{p}} \\
U_{\mathrm{f}}
\end{bmatrix}=\mathcal{H}_{L}(U_{\mathrm{d}}),
\begin{bmatrix}
E_{\mathrm{p}} \\
E_{\mathrm{f}}
\end{bmatrix}=\mathcal{H}_{L}(E_{\mathrm{d}}),
\begin{bmatrix}
X_{\mathrm{p}} \\
X_{\mathrm{f}}
\end{bmatrix}=\mathcal{H}_{L}(X_{\mathrm{d}}),
\end{equation}
where $L=T_{\mathrm{ini}}+N$, $U_{\mathrm{p}} $ and $ U_{\mathrm{f}} $ contain the upper $T_{\mathrm{ini}}$ rows and lower $ N $ rows of $\mathcal{H}_{L}(U_{\mathrm{d}})$, respectively (similarly for $E_{\mathrm{p}}$ and $E_{\mathrm{f}}$, $X_{\mathrm{p}}$ and $X_{\mathrm{f}}$).

Note that for the persistently excitation requirement of order $T_{\mathrm{ini}}+N$, it must hold that $ T \geq 2\left(T_{\mathrm{ini}}+N+2n\right)-1 $~\cite{Coulson2019data,Willems2005note}. This necessary condition indicates that the column number should be larger than the row number in Hankel matrices. 

\subsection{Data-Driven Reachable Set Computation}
\label{Sec:3B}
In this paper, we focus on controlling mixed platoons under process noise and external disturbances. To ensure the robustness of the control system, the reachable set technique is introduced.

\textit{1) System Over-Approximation:} For the mixed platoon system~\eqref{Eq:DiscreteSystem}, multiple models $\begin{bmatrix}A & B & H\end{bmatrix}$ consistent with collected data are considered due to the presence of process noise $w(k)$. Here, we utilize collected data to construct the matrix zonotope set $\mathcal{M}_{ABH}$ to over-approximate all possible system models consistent with the noisy data, as shown below.
\begin{lemma}      
	\label{Lemma:M_ABH}
	Given the data sequences $U_{-}$, $E_{-}$, $X_{-}$, and $X_{+}$ collected from the mixed platoon system~\eqref{Eq:DiscreteSystem}, assume the process noise $w(k)$ and disturbance $\epsilon(k)$ are bounded by zonotope sets, given by:
	\begin{equation}
	\label{Eq:W_Zonotope}
	w(k) \in \mathcal{Z}_w, \quad \epsilon(k) \in \mathcal{Z}_{\epsilon}.
	\end{equation}
	If the matrix $ \begin{bmatrix} X_{-}^{\top} & U_{-}^{\top} & E_{-}^{\top}\end{bmatrix}$ is of full row rank, then the set of all possible $\begin{bmatrix}A & B & H\end{bmatrix}$ can be obtained by:
	\begin{equation}
	\label{Eq:M_ABH}
	\mathcal{M}_{ABH}=\left(X_{+}-\mathcal{M}_w\right)\begin{bmatrix}
	X_{-} \\
	U_{-} \\
	E_{-}
	\end{bmatrix}^{\dagger},
	\end{equation}
which provides a data-driven over-approximation of system dynamics. In~\eqref{Eq:M_ABH}, $\dagger$ denotes Moore–Penrose pseudoinverse, and $\mathcal{M}_w = \left \langle C_{\mathcal{M}_w}, \left[G_{\mathcal{M}_w}^{(1)}, G_{\mathcal{M}_w}^{(2)},\ldots, G_{\mathcal{M}_w}^{(\gamma_{\mathcal{M}_w})}\right]\right \rangle$ is a matrix zonotope set resulting from the noise zonotope set $\mathcal{Z}_w$, with $\gamma_{\mathcal{M}_w} \in \mathbb{N}$ is the number of generator matrices.
\end{lemma}

\textit{Proof:} For the mixed platoon system~\eqref{Eq:DiscreteSystem}, we have
\begin{equation}
X_{+} = \begin{bmatrix}A & B & H\end{bmatrix}\begin{bmatrix}
X_{-} \\
U_{-} \\
E_{-}
\end{bmatrix} + W_{-}.
\end{equation}
Since $ \begin{bmatrix} X_{-}^{\top} & U_{-}^{\top} & E_{-}^{\top}\end{bmatrix}$ is of full row rank, we have
\begin{equation}
\begin{bmatrix}A & B & H\end{bmatrix} =\left(X_{+}-W_{-}\right)\begin{bmatrix}
X_{-} \\
U_{-} \\
E_{-}
\end{bmatrix}^{\dagger}.
\end{equation}
This allows us to use the bound $\mathcal{M}_w$ to obtain~\eqref{Eq:M_ABH}. Thus, the matrix zonotope $\mathcal{M}_{ABH}$ serves as an over-approximated set for system models $\begin{bmatrix}A & B & H\end{bmatrix}$, given the existence of process noise. $\hfill\blacksquare$ 

\vspace{0.2em}
\textit{2) System Decomposition:} Next, the system described in~\eqref{Eq:DiscreteSystem} is decoupled into two parts: nominal dynamics and error dynamics, denoted as follows:
\begin{subequations}
	\label{Eq:DynamicsDecomposition}
	\begin{equation}
	\label{Eq:NominalDynamics}
	{x}_\mathrm{z}(k+1)=A{x}_\mathrm{z}(k)+B{u}_\mathrm{z}(k)+H\epsilon_\mathrm{z}(k),
	\end{equation}
	\begin{equation}
	\label{Eq:ErrorDynamics}
	{x}_\mathrm{e}(k+1)=A{x}_\mathrm{e}(k)+Bu_\mathrm{e}(k)+H\epsilon_\mathrm{e}(k)+ w(k),
	\end{equation}
\end{subequations}
where ${x}_\mathrm{z}(k)$, $u_\mathrm{z}(k)$, $\epsilon_\mathrm{z}(k)$ and ${x}_\mathrm{e}(k)$, $u_\mathrm{e}(k)$, $\epsilon_\mathrm{e}(k)$ represent the state, control input, and disturbance input of the nominal dynamics system and error dynamics system, respectively. Precisely, we have
\begin{equation}
\label{Eq:DecoupledSystem}
\begin{cases}
{x}(k) ={x}_\mathrm{z}(k) + {x}_\mathrm{e}(k),	 \\
{u}(k) ={u}_\mathrm{z}(k) + {u}_\mathrm{e}(k), \\
{\epsilon}(k) ={\epsilon}_\mathrm{z}(k) + {\epsilon}_\mathrm{e}(k),
\end{cases}
\end{equation}
with
\begin{equation}
\label{Eq:DisturbanceDecouple}
{\epsilon}_\mathrm{z}(k) = 0 ,\quad {\epsilon}_\mathrm{e}(k) = {\epsilon}.
\end{equation}
This decomposition indicates that the noise and disturbance are only considered in the error system, not the nominal one. 
Recall that $A$, $B$, and $H$ in~\eqref{Eq:DynamicsDecomposition} are unknown but belong to $\mathcal{M}_{ABH}$.

\vspace{0.2em}
\textit{3) Feedback Control Law for Error Systems:} For the error system~\eqref{Eq:ErrorDynamics}, we need to design a feedback control law $K \in \mathbb{R}^{1 \times 2n}$ that stabilizes all possible $ A $ and $ B $. We first compute a set $\mathcal{M}_{AB}$ that includes all possible $\begin{bmatrix}A & B\end{bmatrix}$, expressed as follows:
\begin{equation}
\mathcal{M}_{AB}=\mathcal{M}_{ABH} \begin{bmatrix}
\mathbb{I}_{2n+1} \\
\mathbb{O}_{1 \times (2n+1)}
\end{bmatrix},
\end{equation}
where $\mathbb{I}$ and $\mathbb{O}$ denote the unit and zero matrices with appropriate dimensions, respectively. Then, we apply random sampling methods to solve $K$ based on~\cite[Lemma 7]{Russo2023tube}. Specifically, we obtain a batch $\mathcal{S} _{N_k} = \{(A^{(1)}, B^{(1)}), (A^{(2)}, B^{(2)}), \ldots, (A^{(N_k)}, B^{(N_k)})\} $ by sampling $\mathcal{M}_{AB}$ in an i.i.d. manner according to a probability measure $\mathbb{P}_{N_K}$. Then, we solve the following linear matrix inequality
\begin{equation}
\begin{bmatrix}
-P & A P+B Z \\
PA^{\top}+Z^{\top}B^{\top} & -P
\end{bmatrix} \prec 0, \quad \forall(A, B) \in \mathcal{S} _{N_k},
\end{equation}
in $ P \succ 0$ and $Z$, which yields 
\begin{equation}
\label{Eq:K}
K =ZP^{-1}.
\end{equation}
Note that $K$ depends on the choice of $\mathcal{S} _{N_k}$. The following lemma provides a robustness guarantee for $K$.

\begin{lemma}[Robustness guarantee for $K$~\cite{Russo2023tube}]      
	\label{Lemma:K}
	For accuracy $\varepsilon \in (0,1)$ and confidence $\delta \in (0,1)$, let $N_k \geq \frac{5}{\varepsilon}\left(\ln \frac{4}{\delta}+d \ln \frac{40}{\varepsilon}\right)$ with $ d=4 n \log _{2}\left(2 e (2n)^{2}(2n+1)\right) $. Assume that $K$ is computed according to $\mathcal{S}_{N_k}$, and satisfied that $\rho\left(A+B K\right) \leq 1$ for every pair $(A,B) \in \mathcal{S}_{N_k}$  Then, with probability at least $ 1-\delta $ we have
	\begin{equation}
	\mathbb{P}\left((A, B) \in \mathcal{M}_{AB}, \rho\left(A+B K\right) \geq 1\right) \leq \varepsilon,
	\end{equation}
	where $\rho $ is the spectral radius, and $\rho\left(A+B K\right) \geq 1$ means that Schur stability is not satisfied. 
\end{lemma}

\vspace{0.2em}
\textit{4) Error State Reachable Sets:} Based on~\eqref{Eq:W_Zonotope},~\eqref{Eq:M_ABH},~\eqref{Eq:ErrorDynamics} and~\eqref{Eq:K}, the recursive relation for the error state reachable set can be obtained as follows:
\begin{equation}
\label{Eq:ReachableSet}
\hat{\mathcal{R}}^\mathrm{e}_{k+i+1}=\mathcal{M}_{A B H}\left(\hat{\mathcal{R}}^\mathrm{e}_{k+i} \times K\hat{\mathcal{R}}^\mathrm{e}_{k+i} \times \mathcal{Z}_{\epsilon}\right)+\mathcal{Z}_w,
\end{equation}
where $\hat{\mathcal{R}}^\mathrm{e}_{k+i}$ is an over-approximated reachable set for the state $ x_\mathrm{e}(k+i)$ of the error dynamics system.

\subsection{\method{RDeeP-LCC} Optimization Formulation}
\label{Sec:3C}

We proceed to design the \method{RDeeP-LCC} optimization formulation. The detailed process is described below:

\textit{1) Trajectory Definition:} For each time step $ k $, we define the state trajectory $ x_\mathrm{ini} $ in the past $ T_\mathrm{ini} $ steps and the predicted state trajectory $x_z$ of the nominal system  in the next $N$ steps, denoted as follows:
\begin{equation}
\label{Eq:StateTrajectory}
\begin{cases}
x_\mathrm{ini} = \mathrm{col}(x(k-T_\mathrm{ini}+1),x(k-T_\mathrm{ini}+2),\ldots,x(k)), \\
x_\mathrm{z} = \mathrm{col}(x_\mathrm{z}(k+1),x_\mathrm{z}(k+2),\ldots,x_\mathrm{z}(k+N)).
\end{cases}
\end{equation}
The control input trajectories $u_\mathrm{ini}$ and $u_\mathrm{z}$, and disturbance input trajectories $\epsilon_{\mathrm{ini}}$ and $\epsilon_\mathrm{z}$ in the past $T_\mathrm{ini}$ steps and future $ N $ steps are defined similarly as in~\eqref{Eq:StateTrajectory}.

\textit{2) Cost Function:} We utilize the quadratic function $ J(x_z,u_z) $ to quantify the control performance by penalizing the states and control inputs, defined as follows:
\begin{equation}
\label{Eq:Cost}
J(x_\mathrm{z},u_\mathrm{z})=\sum_{i=1}^{N}\left(\|x_\mathrm{z}(k+i)\|_{Q}^{2}+\|u_\mathrm{z}(k+i)\|_{R}^{2}\right),
\end{equation}	
where $ Q = \operatorname{diag}\left(\rho_{s}, \rho_{v}, \ldots, \rho_{s}, \rho_{v}\right) \in \mathbb{R}^{2n \times 2n}, R\in \mathbb{R}$ denote weight matrices, with $\rho_{s} $ and $\rho_{v}$ denoting the penalties for spacing deviation and velocity deviation, respectively.

\textit{3) Data-Driven Dynamics:} By Lemma~\ref{Lemma:Fundamental} and~\cite[Proposition 2]{Wang2023deep}, the data-driven dynamics can be given by
\begin{equation}
\label{Eq:WilliemExpanding}
\begin{bmatrix}
X_{\mathrm{p}} \\
U_{\mathrm{p}} \\
E_{\mathrm{p}} \\
X_{\mathrm{f}} \\
U_{\mathrm{f}} \\
E_{\mathrm{f}}
\end{bmatrix} g=\begin{bmatrix}
x_{\mathrm {ini}} \\
u_{\mathrm {ini}} \\
\epsilon_{\mathrm{ini}} \\
x_\mathrm{z} \\
u_\mathrm{z} \\
\epsilon_\mathrm{z}
\end{bmatrix}.
\end{equation}
The existence of $ g \in \mathbb{R}^{T-T_\mathrm{ini}-N+1} $ satisfying~\eqref{Eq:WilliemExpanding} implies that $ x_\mathrm{z} $, $ u_\mathrm{z} $, and $\epsilon_\mathrm{z}$ form a future trajectory of length $N$. Note that to ensure the uniqueness of the future trajectory $ x_z $ for given $x_{\mathrm {ini}}$, $u_{\mathrm {ini }}$, $\epsilon_{\mathrm{ini}}$, $ u_\mathrm{z} $, $\epsilon_\mathrm{z}$, it is required that $ T_\mathrm{ini} \geq 2n $~\cite{Wang2023deep}.

\textit{4) Constraints:} The safety of the mixed platoon system is ensured by imposing the following constraints: 
\begin{equation}
\label{Eq:Constrain}
\begin{cases}
x(k+i) \in \mathcal{X},  \\
u(k+i) \in \mathcal{U},
\end{cases}
\end{equation}
where $\mathcal{X}=\left\{x(k) \in \mathbb{R}^{2n}\mid|x(k)|\leq \mathbf{1}_{n} \otimes x_{\max }\right\}$ is the state constraint, with $x_{\max }=\left[\tilde{s}_{\max }, \tilde{v}_{\max }\right]^{\top} $, where $\tilde{s}_{\max }$ and $\tilde{v}_{\max }$ are the constraint limits for spacing deviation and velocity deviation, respectively. In the control input constraint $\mathcal{U}=\left\{u(k) \in \mathbb{R}\mid|u(k)|\leq u_{\max }\right\}$, $u_{\max }$ denotes the maximum control input for the CAV.

Combining~\eqref{Eq:ReachableSet} and~\eqref{Eq:Constrain}, the constraint set for the nominal system~\eqref{Eq:NominalDynamics} can be calculated as follows:
\begin{equation}
\label{Eq:RConstraint}
\begin{cases}
\mathcal{X}_\mathrm{z}(k+i)= \mathcal{X} - \hat{\mathcal{R}}_{k+i}^\mathrm{e},\\
\mathcal{U}_\mathrm{z}(k+i)= \mathcal{U} - K\hat{\mathcal{R}}_{k+i}^\mathrm{e},
\end{cases}
\end{equation}
which yields the constraints for the predicted trajectory of the nominal system, given by:
\begin{equation}
\label{Eq:NormalConstraint}
\begin{cases}
x_\mathrm{z}(k+i) \in \mathcal{X}_\mathrm{z}(k+i),\\
u_\mathrm{z}(k+i) \in \mathcal{U}_\mathrm{z}(k+i).
\end{cases}
\end{equation}

For the future disturbance sequence $\epsilon_z$ of the nominal system, recalling~\eqref{Eq:DisturbanceDecouple}, we have
\begin{equation}
\label{Eq:DisturbanceConstraint}
\epsilon_\mathrm{z}(k+i) = 0.
\end{equation}


\textit{5) \method{RDeeP-LCC} Optimization Formulation:} The final optimization problem is formulated as follows to solve the control input for the CAVs:
\begin{equation}
\label{Eq:OptimizationProblemFinal}
\begin{array}{l}
\min\limits_{g, u_z, x_z,\sigma} \ J(x_z, u_z) + \lambda_\mathrm{g}\|g\|_{2}^{2}+\lambda_\mathrm{\sigma}\|\sigma\|_{2}^{2} \\
\mathrm{ s.t. } 
\begin{array}{l}
\begin{bmatrix}
X_{\mathrm{p}} \\
U_{\mathrm{p}} \\
E_{\mathrm{p}} \\
X_{\mathrm{f}} \\
U_{\mathrm{f}} \\
E_{\mathrm{f}}
\end{bmatrix} g=\begin{bmatrix}
x_{\mathrm {ini}} \\
u_{\mathrm {ini}} \\
\epsilon_{\mathrm{ini}} \\
x_\mathrm{z} \\
u_\mathrm{z} \\
\epsilon_\mathrm{z}
\end{bmatrix}+
\begin{bmatrix}
\sigma \\
0\\
0 \\
0 \\
0 \\
0
\end{bmatrix},\\
\eqref{Eq:NormalConstraint},~\eqref{Eq:DisturbanceConstraint},
\end{array}
\end{array}
\end{equation}
Motivated by the literature~\cite{Coulson2019data,Wang2023deep}, we ensure the feasibility of the optimization problem~\eqref{Eq:OptimizationProblemFinal} by penalizing $ g $ with regularization and introducing a slack variable $ \sigma \in \mathbb{R}^{2nT_\mathrm{ini}} $ in~\eqref{Eq:WilliemExpanding},
where $\lambda_{g},\lambda_{\sigma} \geq 0 $ denote the regularization coefficients. This regularization is necessary since the real mixed platoon system has nonlinear dynamics, while the data-driven dynamics~\eqref{Eq:WilliemExpanding} is only applicable to LTI systems.

Solving~\eqref{Eq:OptimizationProblemFinal} yields an optimal control sequence $ u_z$ and the predicted state sequence $x_z$ of the nominal system. Then, by
\begin{equation}
\begin{aligned}
u(k+1) &= u_\mathrm{z}(k+1) + u_\mathrm{e}(k+1) 
\\&= u_\mathrm{z}(k+1) + K({x}(k+1)-{x}_\mathrm{z}(k+1)),
\end{aligned}
\end{equation}
where $K$ is calculated from~\eqref{Eq:K}, we obtain the final control input for the CAV.

\begin{remark}
	It is worth noting that in~\eqref{Eq:OptimizationProblemFinal}, $x_{\mathrm {ini}},u_{\mathrm {ini}},\epsilon_{\mathrm {ini}}$ represent the past trajectories of the actual system~\eqref{Eq:DiscreteSystem}, while $x_\mathrm{z}, u_\mathrm{z},\epsilon_\mathrm{z}$ denote the predicted trajectories of the nominal system~\eqref{Eq:NominalDynamics}. As shown in~\eqref{Eq:DisturbanceDecouple}, we assume $\epsilon_\mathrm{z} = 0$, and capture the actual disturbance $\epsilon_\mathrm{e} = \epsilon$ solely in the error system~\eqref{Eq:ErrorDynamics}. Provided $\epsilon(k) \in \mathcal{Z}_{\epsilon}$ in~\eqref{Eq:W_Zonotope}, the disturbance effect is further incorporated into the calculation of the error reachable set~\eqref{Eq:ReachableSet}, resulting in a more stringent constraint~\eqref{Eq:NormalConstraint} on $x_\mathrm{z}$ and $u_\mathrm{z}$ of the nominal system. This approach addresses the influence of process noise and unknown future disturbances, which are neglected in the standard \method{DeeP-LCC}~\cite{Wang2023deep}.
\end{remark}

\section{Simulation Results}
\label{Sec:4}	
In this section, we carry out nonlinear traffic simulations to verify the performance of the proposed \method{RDeeP-LCC} method.

\subsection{Simulation Setup}
For the mixed platoon system depicted in Fig.~\ref{Fig:MixedPlatoon}, we set the platoon size as $ n = 3 $. The dynamics of vehicles are described in~\eqref{Eq:DynamicsModel}. For HDVs, we utilize the OVM model~\cite{Bando1995dynamical} to describe its car-following behaviors, with the specific model and the parameter selection presented in~\cite{wang2020controllability}. The parameters of the \method{RDeeP-LCC} in the simulation are set as follows: 

\begin{itemize}
	\item 
	For offline data collection, near the equilibrium state $ v^{*}=\SI{15}{m/s} $, we utilize a random control input within $\left[-0.2, 0.2\right]$ for the CAVs and apply a random disturbance within $\left[-0.5, 0.5\right]$ to the head vehicle's velocity. Then the offline pre-collected trajectories of length $ T = 1000 $ with a sampling interval of $\SI{0.1}{s}$ are used to construct~\eqref{Eq:Xsequence}-\eqref{Eq:Hankel}. Based on the pre-collected data sequences, $\mathcal{M}_{ABH}$ can be obtained using~\eqref{Eq:M_ABH}, and $K$ is computed using Lemma~\ref{Lemma:K} with $\varepsilon = 0.01$ and $\delta = 0.001$. 
	\item 
	For online predictive control, the past sequence length $ T_\mathrm{ini}= 20 $ and the future sequence length $ N=5 $ are selected in~\eqref{Eq:StateTrajectory}. For the cost function~\eqref{Eq:Cost}, the coefficients are set to $\rho_{s}=0.5$, $\rho_{v}=1$, and $R=0.1$. We set $x_{\max }=\left[7, 7\right]^{\top} $ and $u_{\max}=5$ in the constraints~\eqref{Eq:Constrain}. In the optimization formulation~\eqref{Eq:OptimizationProblemFinal}, the parameters are set to $\lambda_\mathrm{g}=10$ and $\lambda_\mathrm{\sigma}=10$. Based on the localization requirements for local roads in the United States~\cite{reid2019localization}, we set the range of bounded process noise is $|w(k)| \leq 0.05$. 
\end{itemize}

In addition, we choose standard MPC with knowledge of system dynamics~\eqref{Eq:DiscreteSystem} and standard \method{DeeP-LCC} with the same pre-collected data as baseline methods. All shared parameters of standard MPC and \method{DeeP-LCC} have the same values as those of \method{RDeeP-LCC}, except for $ N=20$ in \method{DeeP-LCC}, which follows the setup in~\cite{Wang2023deep}.

In order to quantify the performance, this paper adopts the velocity mean absolute deviation $ R_\mathrm{m} $ and velocity standard deviation $ R_\mathrm{s} $ as performance indices, given by:
\begin{subequations}
	\begin{equation}
	\label{Eq:MADvalue}
	R_\mathrm{m}=\frac{1}{T_\mathrm{s}} \frac{1}{n} \sum_{t=0}^{T_\mathrm{s}} \sum_{i=1}^{n}\left|v_{i}(t)-v^{*}\right|,
	\end{equation}
	\begin{equation}
	\label{Eq:SDvalue}
	R_\mathrm{s}=\sqrt{\frac{1}{T_\mathrm{s}} \frac{1}{n} \sum_{t=0}^{T_\mathrm{s}} \sum_{i=1}^{n}\left(v_{i}(t)-v^{*}\right)^{2}},
	\end{equation}	
\end{subequations}
where $T_\mathrm{s}$ is the total simulation time.

\subsection{Simulation Results}
\begin{figure}[t]
	\vspace{1mm}
	\centering
	\subfigure[All HDVs]{\includegraphics[width=4.2cm]{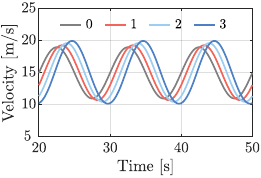}}
	\subfigure[Standard MPC]{\includegraphics[width=4.2cm]{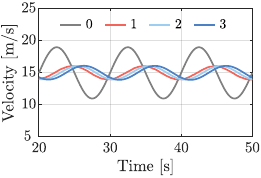}}\\
	\vspace{-3mm}
	\subfigure[Standard \method{DeeP-LCC}]{\includegraphics[width=4.2cm]{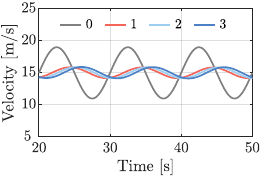}}
	\subfigure[\method{RDeeP-LCC}]{\includegraphics[width=4.2cm]{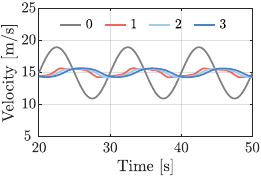}}\\
	\vspace{-2mm}
	\caption{Velocity profiles under different control methods for Simulation A. The gray profile, the red profile, and the blue profiles represent the head vehicle (indexed 0), the CAV (indexed 1), and the HDVs (indexed 2, 3), respectively. Note that \method{DeeP-LCC} and \method{RDeeP-LCC} use the same dataset.}
	\label{Fig:SimulationResults}
	\vspace{-3mm}
\end{figure}

\textit{1) Simulation A (Near Equilibrium State):} To reproduce a traffic wave scenario, inspired by~\cite{jin2016optimal,Wang2023deep}, we introduce a sinusoidal disturbance with an amplitude of $\SI{4}{m/s}$ and a period of $\SI{10}{s}$ on the head vehicle. The velocity profiles of our proposed approach and other baseline methods are shown in Fig.~\ref{Fig:SimulationResults}. From Fig.~\ref{Fig:SimulationResults}(a), it is evident that the perturbation is amplified during propagation when all vehicles are HDVs. Conversely, when CAVs utilize standard MPC, standard \method{DeeP-LCC}, or \method{RDeeP-LCC}, as shown in Fig.~\ref{Fig:SimulationResults}(b)-(d), respectively, the magnitude of the perturbation is notably reduced. This demonstrates the capability of the CAVs to mitigate undesired disturbances when employing either of the three methods. 

Note that different pre-collected data may affect the data-driven controller's performance due to the existence of process noise. To further qualitatively compare the performance, we collect $20$ random data sets with a consistent length of $ T = 1000 $ to construct the Hankel matrices~\eqref{Eq:Hankel}. We then conduct simulations using these data sets and analyze the performance indices \eqref{Eq:MADvalue} and \eqref{Eq:SDvalue}. The results, presented in Table~\ref{tab:experiment_A}, indicate that our \method{RDeeP-LCC} approach exhibits the smallest mean values of \eqref{Eq:MADvalue} and \eqref{Eq:SDvalue} compared to the baseline methods. Specifically, compared to the cases of all HDVs, our approach achieves average reductions of $\SI{79.9}{\%}$ and $\SI{81.7}{\%}$, while the baseline methods achieve less reductions. This reveals the superior performance of our approach in suppressing undesired traffic oscillations, as higher values of $ R_\mathrm{m} $ and $ R_\mathrm{s} $ indicate a larger velocity fluctuation and a weaker wave-dampening ability of the CAVs' controller. 

\begin{table}[t]
	\centering
	\renewcommand{\arraystretch}{1.4}
	\caption{The $ R_\mathrm{m}$ and $ R_\mathrm{s}$ in Simulation A}
	\label{tab:experiment_A}
	\begin{tabular}{m{0.7cm}<{\centering}|m{1.36cm}<{\centering}|m{1.36cm}<{\centering}|m{1.36cm}<{\centering}|m{1.6cm}<{\centering}}
		\hline
		Index    &All HDVs &MPC &\method{DeeP-LCC} &\method{RDeeP-LCC}\\ \hline
		\multirow{2}{*}{$R_\mathrm{m}$}    &2.412 &0.673  &0.511 &0.486  \\ 
		&$(--)$  &$(\downarrow \SI{72.1}{\%})$  &$(\downarrow \SI{78.8}{\%})$  &$(\downarrow \SI{79.9}{\%})$  \\ \hline
		\multirow{2}{*}{$R_\mathrm{s}$} &2.892 &0.747  &0.553 &0.530 \\ 
		&$(--)$  &$(\downarrow \SI{74.2}{\%})$  &$(\downarrow \SI{80.9}{\%})$  &$(\downarrow \SI{81.7}{\%})$  \\ \hline
	\end{tabular}
	\vspace{-1mm}
\end{table}

\textit{2) Simulation B (Standard Test Cycle):} Inspired by the experiments conducted in~\cite{Wang2023deep}, we devise a comprehensive acceleration and deceleration scenario based on the New European Driving Cycle (NEDC) to assess the effectiveness of the proposed \method{RDeeP-LCC} in enhancing platoon performance. In this scenario, We adopt the ECE Elementary Urban Cycle (Part One of the Urban Driving Cycle) as a trajectory for the head vehicle. It is noteworthy that, in simulation B, we assume the real-time velocity of the head vehicle as the equilibrium velocity for online predictive control.

\begin{figure*}[t]
	\vspace{1mm}
	\centering
	\subfigure[All HDVs]{\includegraphics[width=8.8cm]{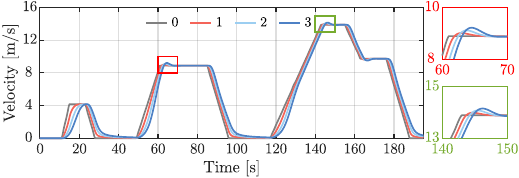}}
	\vspace{-3mm}
	\subfigure[Standard MPC]{\includegraphics[width=8.8cm]{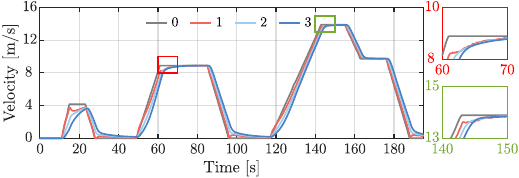}}\\
	\vspace{-0mm}
	\subfigure[Standard \method{DeeP-LCC}]{\includegraphics[width=8.8cm]{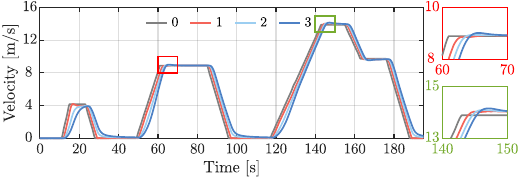}}
	\vspace{-3mm}
	\subfigure[\method{RDeeP-LCC}]{\includegraphics[width=8.8cm]{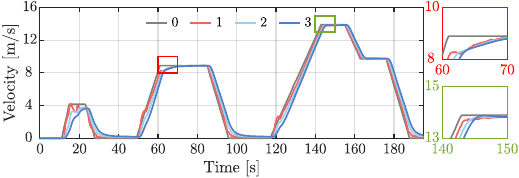}}\\
	\vspace{0.5mm}
	\caption{Velocity profiles under different control methods for Simulation B. The graphs are colored corresponding to the color of profiles in Fig.~\ref{Fig:SimulationResults}.}
	\label{Fig:SimulationNEDCResults}
	\vspace{-3mm}
\end{figure*}

The simulation results depicted in Fig.~\ref{Fig:SimulationNEDCResults} reveal the performance of mixed platoons using different control methods. Overall, the platoons successfully track the desired trajectory for all the three methods. However, there are instances, such as at about $\SI{65}{s}$ and $\SI{145}{s}$, where both All HDVs and \method{DeeP-LCC} exhibit noticeable overshoot, as depicted in the insets of Fig.~\ref{Fig:SimulationNEDCResults}(a),(c). In contrast, both standard MPC and \method{RDeeP-LCC} effectively mitigate overshoot, resulting in superior tracking performance in these cases, as illustrated in the insets of Fig.~\ref{Fig:SimulationNEDCResults}(b),(d). This highlights the robustness and effectiveness of \method{RDeeP-LCC} in achieving precise control under time-varying traffic conditions.

Table~\ref{tab:experiment_B} illustrates the $R_\mathrm{m}$ and $R_\mathrm{s}$ indices, which provide straightforward insights into the control performance in terms of velocity errors. It is observed that the proposed \method{RDeeP-LCC} method shows the best performance compared to all other cases. It is worth noting that the standard \method{DeeP-LCC} shows the highest values for $R_\mathrm{m}$ and $R_\mathrm{s}$. This means that under the influence of process noise and disturbances, the data-driven predictive control methods without explicit robust design could have unsatisfactory tracking performance, which could be even worse than HDV's natural behaviors in traffic flow with time-varying equilibrium states. In contrast, our proposed method enhances robustness against process noise and disturbances for mixed platoon control. 


\begin{table}[]
\vspace{2mm}
	\centering
	\renewcommand{\arraystretch}{1.4}
	\caption{The $ R_\mathrm{m}$ and $ R_\mathrm{s}$ in Simulation B}
	\label{tab:experiment_B}
	\begin{tabular}{m{0.7cm}<{\centering}|m{1.36cm}<{\centering}|m{1.36cm}<{\centering}|m{1.36cm}<{\centering}|m{1.6cm}<{\centering}}
		\hline
		Index    &All HDVs &MPC &\method{DeeP-LCC} &\method{RDeeP-LCC}\\ \hline
		\multirow{2}{*}{$R_\mathrm{m}$}    &0.566 &0.537  &0.620 &0.526  \\ 
		&$(--)$  &$(\downarrow \SI{5.1}{\%})$  &$(\uparrow \SI{9.5}{\%})$  &$(\downarrow \SI{7.1}{\%})$  \\ \hline
		\multirow{2}{*}{$R_\mathrm{s}$} &0.889 &0.785  &0.948 &0.778 \\ 
		&$(--)$  &$(\downarrow \SI{11.7}{\%})$  &$(\uparrow \SI{6.6}{\%})$  &$(\downarrow \SI{12.5}{\%})$  \\ \hline
	\end{tabular}
	\vspace{-3mm}
\end{table}

\section{Conclusions}
\label{Sec:5}
In this paper, we have proposed \method{RDeeP-LCC} for mixed platoon control to address the existence of noise and disturbances by combining data-driven reachable set analysis and the standard data-enabled predictive control framework. Traffic simulation results have demonstrated that our method effectively reduces undesired traffic waves and achieves excellent tracking performance compared to all HDVs, standard MPC, and standard \method{DeeP-LCC}. Future directions include investigating mixed platoon control under communication delays and experimental validations with real human drivers in the loop. 


\bibliographystyle{IEEEtran}
\bibliography{IEEEabrv,Reference}

\end{document}